\documentclass[reprint,aps,prl,amsmath,amssymb,superscriptaddress]{revtex4-2}
\usepackage[utf8]{inputenc}
\usepackage{graphicx}
\usepackage{bm}
\usepackage{physics}
\usepackage{xcolor}
\usepackage[colorlinks=true,allcolors=blue]{hyperref}
\graphicspath{{figs/}}
\begin{document}

\title{Chiral $p$-Wave Superconductivity Proximate to a Fractional Chern
Insulator and a Topological Charge Density Wave in a Moir\'e Chern Band}
\author{Yunze Ouyang}
\affiliation{School of Physics and Technology, Wuhan University, Wuhan 430072, China}
\author{Fengcheng Wu}
\email[]{wufcheng@whu.edu.cn}
\affiliation{School of Physics and Technology, Wuhan University, Wuhan 430072, China}
\affiliation{Wuhan Institute of Quantum Technology, Wuhan 430206, China}

\begin{abstract}
We study a spin-polarized moir\'e Chern band of twisted
MoTe$_2$ with short-range repulsion, finding superconductivity
proximate to a fractional Chern insulator at filling $\nu=2/3$
and a topological charge density wave at $\nu=3/4$.
Exact diagonalization reveals pair binding over
$0.7\lesssim\nu\lesssim0.77$, consistent with experiment,
along with finite superfluid stiffness, chiral $p$-wave
pairing correlations, and adiabatic connectivity to an attractive model.
At weak repulsion, Kohn--Luttinger theory supports a
Fermi-surface instability with the same pairing symmetry.
The superconducting state has a
Bogoliubov--de Gennes Chern number $C_{\mathrm{BdG}}=-1$,
opposite to the normal-state Chern number at $\nu=1$.
Our results reveal interplay among superconductivity,
fractional topology, and charge order.
\end{abstract}

\maketitle

\textit{Introduction.}---
Topological superconductors can host exotic excitations, including
Majorana modes, and offer a promising avenue for exploring
topological quantum phenomena \cite{Qi2011Top,Nayak2008Non}.
Two-dimensional materials and their moir\'e heterostructures provide
a potential platform for realizing such phases through the interplay
of quantum geometry and strong correlations.
Recent experiments in rhombohedral multilayer graphene have reported signatures
of chiral superconductivity emerging from spin- and valley-polarized metals \cite{Han2025signatures}.
Twisted bilayer MoTe$_2$ (tMoTe$_2$) offers another rich platform,
hosting fractional Chern insulators (FCIs) \cite{Cai2023signatures,Zeng2023thermodynamic,Park2023observation,Xu2023observation}, nearby superconductivity (SC) \cite{Xu2025signatures},
and reentrant integer quantum anomalous Hall states,
possibly associated with charge order \cite{Xu2025signatures}, within a spin-polarized (spin-valley-locked) moir\'e Chern band.
The proximity of these phases raises fundamental questions
about their microscopic origins and interplay.

Several mechanisms have been proposed for SC near FCIs \cite{Shi2025doping,Pichler2026microscopic,Nosov2026anyon,Xu2025chiral,Guerci2025from,Guerci2026topological, Wang2025chiral, Hu2026quantum}.
One scenario involves the condensation of doped anyons
\cite{Laughlin1988SC,Laughlin1988relationship,Shi2025doping,Pichler2026microscopic,Nosov2026anyon},
while another invokes pairing instability of a Fermi liquid driven  by repulsion, such as in the Kohn--Luttinger (KL) mechanism \cite{Kohn1965new, Xu2025chiral,Shavit2025quantum,Jahin2026enhanced}.
Recent numerical studies have found evidence of chiral SC
near an FCI in a flat Chern band with generalized lowest-Landau-level (LLL)
wave functions \cite{Guerci2025from,Guerci2026topological} and in an LLL
subject to a periodic potential \cite{Wang2025chiral}.
In the latter, superconductivity and charge-density-wave order
closely compete, making their relative stability subtle \cite{Wang2025chiral}.
Understanding how repulsive interactions give rise to SC
and organize competing phases in generic Chern
bands is a compelling challenge with broad implications for
strongly correlated quantum matter.

In this Letter, we show that short-range repulsion in a
spin-polarized moir\'e Chern band of tMoTe$_2$ gives rise to
chiral SC proximate to an FCI at $\nu=2/3$
and a topological charge density wave (TCDW) at $\nu=3/4$.
Large-scale exact diagonalization (ED) reveals signatures of SC over
$0.7\lesssim\nu\lesssim0.77$, consistent with experiment
\cite{Xu2025signatures}, while KL theory at weak interaction supports a
repulsion-driven Fermi-surface instability with pairing
symmetry consistent with ED.
The resulting superconducting state is topological with $C_{\mathrm{BdG}}=-1$.
Notably, the same interaction strength can support the FCI
and SC at nearby fillings, while stronger repulsion favors
the TCDW. Our results resolve the interplay among superconductivity,
fractional topology, and charge order in a Chern band with nonideal quantum geometry and finite bandwidth.

\textit{Model.}---
Since the SC emerges from a spin-polarized metal in tMoTe$_2$, we assume full spin polarization. The continuum model for spin-up states \cite{Wu2019topological}, after a particle-hole transformation, reads
\begin{equation}
\mathcal{H} = -\begin{pmatrix}
h_+(\vb*{r}) + \Delta_+(\vb*{r}) & \Delta_T^\dagger(\vb*{r}) \\
\Delta_T(\vb*{r}) & h_-(\vb*{r}) + \Delta_-(\vb*{r})
\end{pmatrix}.
\end{equation}
The kinetic term is $h_l(\vb*{r}) = -\hbar^2(-i\partial_{\vb*{r}} + \vb*{\kappa}_l)^2/2m^*$, where $l = +1$ ($-1$) labels the bottom (top) layer and $\vb*{\kappa}_{\pm}$ are  the moir\'e Brillouin zone (BZ) corners. The intralayer potential and interlayer tunneling are $\Delta_l(\vb*{r}) = 2v\sum_{j=1,3,5}\cos(\vb*{g}_j\cdot\vb*{r} + l\phi)$ and  $\Delta_T(\vb*{r}) = w\,(1 + e^{-i\vb*{g}_2\cdot\vb*{r}} + e^{-i\vb*{g}_3\cdot\vb*{r}})$, respectively. Here $\vb*{g}_j = (4\pi/\sqrt{3}a_M)\,[\cos(\pi(j-1)/3), \sin(\pi(j-1)/3)]$ are the  moir\'e reciprocal lattice vectors, where $a_M = a_0/\theta$ is the moir\'e lattice constant, $a_0$ is the monolayer lattice constant, and $\theta$ is the twist angle. We adopt the parameters $m^* = 0.6\,m_e$, $v = 20.8$~meV, $\phi = 107.7^\circ$, $w = -23.8$~meV, and $a_0 = 3.52$~\AA~\cite{Wang2024fractional}, where $m_e$ is the bare electron mass.  We fix $\theta=3.83^\circ$, guided by the experiment \cite{Xu2025signatures}.

\begin{figure}
\centering
\includegraphics[width=\columnwidth]{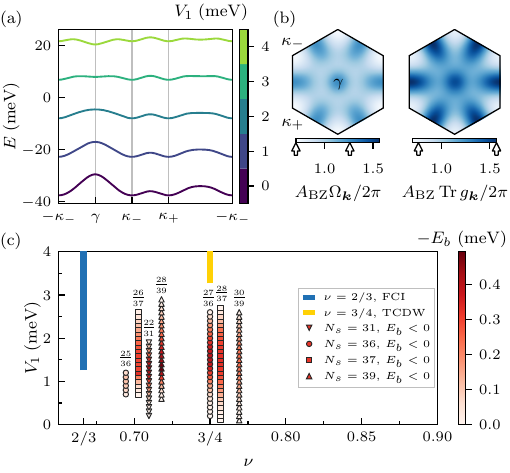}
\caption{(a) Hartree--Fock band at filling $\nu = 1$ along high-symmetry lines of the moir\'e BZ for $V_1 = 0$, 1, 2, 3, and 4~meV. (b)  $\Omega_{\vb*{k}}$ (left) and $\Tr g_{\vb*{k}}$ (right) in units of $2\pi/A_{\mathrm{BZ}}$, with $A_{\mathrm{BZ}}$ the BZ area. Arrows mark the data range of each map. (c) ED phase diagram as functions of $\nu$ and $V_1$. Markers, colored by $-E_b$, indicate pair binding for $N_s = 31$ (down triangles), 36 (circles), 37 (squares), and 39 (up triangles); fractions indicate $N/N_s$. Blue and yellow bars mark the FCI and TCDW, respectively, for $N_s=36$.}
\label{fig:1}
\end{figure}

We consider the short-range repulsive interaction $V(\vb*{r}) = V_1 a_M^4 \nabla^2 \delta(\vb*{r})$ with $V_1 > 0$, which is analogous to the LLL Haldane $V_1$ pseudopotential \cite{Haldane1983fractional} and motivated by recent studies \cite{Guerci2025from,Guerci2026topological,Wang2025chiral}. Projecting onto the lowest-energy band of $\mathcal{H}$ yields the many-body Hamiltonian,
\begin{equation}
H = \sum_{\vb*{k}} \varepsilon_{\vb*{k}} c^\dagger_{\vb*{k}} c_{\vb*{k}} + \frac{1}{2A} \sum_{\vb*{q}} V(\vb*{q})\, {:}\bar{n}(\vb*{q})\, \bar{n}(-\vb*{q}){:},
\end{equation}
where the first term $H_0$ describes the single-particle band dispersion $\varepsilon_{\vb*{k}}$, and the second term $H_{1}$ the interaction. Here $c^\dagger_{\vb*{k}}$ creates a particle in the Bloch state $\ket{\vb*{k}}$ of the band, and $\bar{n}(\vb*{q}) = \sum_{\vb*{k}} \mel{\vb*{k}}{e^{-i \vb*{q}\cdot\vb*{r}}}{\vb*{k}+\vb*{q}}\, c^\dagger_{\vb*{k}} c_{\vb*{k}+\vb*{q}}$ denotes the projected density operator. The sum over $\vb*{q}$ runs over all wave vectors, with $V(\vb*{q}) = -V_1 a_M^4 \abs{\vb*{q}}^2$ the Fourier transform of $V(\vb*{r})$. A system of $N_s$ moir\'e unit cells has the area $A = N_s \sqrt{3} a_M^2/2$. 
The filling factor is $\nu = N/N_s$, where $N = \sum_{\vb*{k} \in \text{BZ} } c^\dagger_{\vb*{k}} c_{\vb*{k}}$ is the number of particles.

\begin{figure}
\centering
\includegraphics[width=\columnwidth]{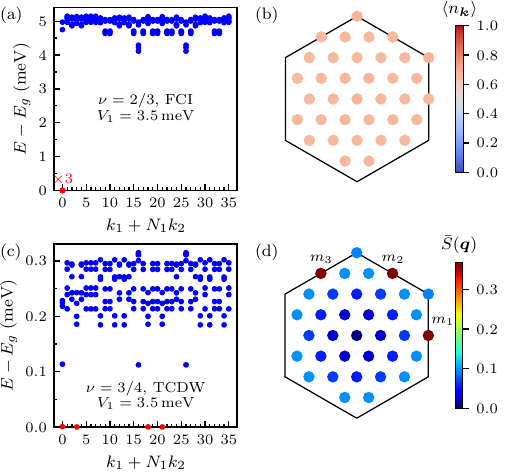}
\caption{(a) ED spectrum of the FCI at $\nu = 2/3$ relative to the ground-state energy $E_g$, with the three quasidegenerate ground states ($\times 3$) in red. (b) $\ev*{n_{\vb*{k}}}$ averaged over the FCI ground states. (c) The TCDW spectrum at $\nu = 3/4$, with four quasidegenerate ground states. (d) Band-projected structure factor $\bar{S}(\vb*{q})$ averaged over the TCDW ground states. All panels are for $N_s = 36$ and $V_1 = 3.5$~meV.}
\label{fig:2}
\end{figure}

We first examine the effective band dispersion at the full filling $\nu=1$, renormalized from the bare dispersion $\varepsilon_{\vb*{k}}$ by Hartree and Fock self-energies.  As $V_1$ increases from $0$ to $4$~meV, the band maximum at the BZ center $\gamma$ evolves into a minimum, and the band becomes nearly flat around $V_1\approx3.5$~meV [Fig.~\ref{fig:1}(a)]. While the interaction renormalizes the dispersion, one-band projection leaves the Bloch states unchanged. Consequently, the band retains its Chern number $C=1$ and quantum geometry. As shown in Fig.~\ref{fig:1}(b), the trace of the quantum metric $\Tr g_{\vb*{k}}$ closely tracks the Berry curvature $\Omega_{\vb*{k}}$ with a comparable magnitude, indicating near-ideal quantum geometry \cite{Wang2021exact,Liu2025theory,Cano2026ideal} and that the Bloch states approximately realize a generalized LLL \cite{Li2025variational,Li2026abelian}. However, the momentum-space inhomogeneity of the quantum geometry leads to interaction-induced modifications of the band dispersion \cite{Liu2025Broken}.

\begin{figure*}
\centering
\includegraphics[width=\textwidth]{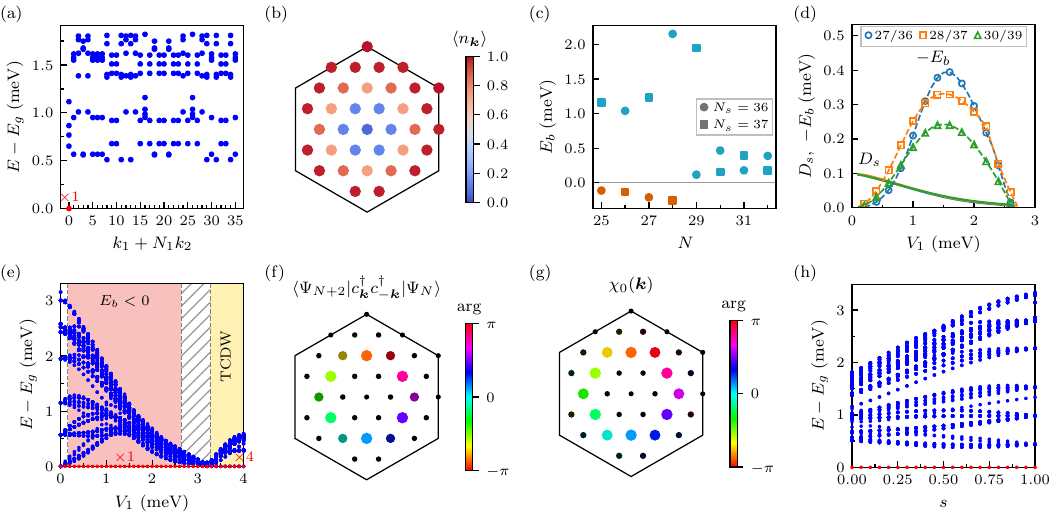}
\caption{(a) ED spectrum, with the nondegenerate ground state ($\times 1$) in red. (b)  $\ev*{n_{\vb*{k}}}$ of the ground state in (a). (c) $E_b(N)$ for $N_s = 36$ (circles) and 37 (squares). Vermilion marks $E_b < 0$ and cyan  $E_b > 0$. (d) $D_s$ (solid lines) and $-E_b$ (dashed lines, open markers) as functions of $V_1$ where $E_b < 0$, for $N/N_s = 27/36$ (circles), $28/37$ (squares), and $30/39$ (triangles). (e) Energy levels as functions of $V_1$, with the ground state in red. Pink, hatched, and yellow shaded regions denote the pair-binding, transition, and TCDW regimes, respectively; $\times n$ indicates an $n$-fold ground-state degeneracy. (f) $F(\vb*{k})$ in the BZ. (g) $\chi_0(\vb*{k})$ for the ground state in (a). 
In (f) and (g), the hue represents the phase, while the brightness and dot size indicate the magnitude. (h) Energy levels along the interpolation from the Hamiltonian in (a) at $s=0$ to the attractive model at $s=1$.
Unless stated otherwise, $N_s = 36$, $N=27$, $\nu = 3/4$, and $V_1 = 1$~meV. In (h), $U=1$~meV.}
\label{fig:3}
\end{figure*}

\textit{Phase diagram.}---
We map out the phase diagram in Fig.~\ref{fig:1}(c) for $2/3\leq\nu\leq1$ and $0\leq V_1\leq4$~meV using large-scale ED on clusters with $N_s=31$, 36, 37, and 39 that preserve sixfold rotational symmetry. At $\nu=2/3$, we identify a robust FCI for $V_1\geq1.26$~meV. Its ED spectrum for $N_s=36$ and $V_1=3.5$~meV exhibits three nearly degenerate ground states separated from excited states by a gap [Fig.~\ref{fig:2}(a)]. The momentum occupation $\ev*{n_{\vb*{k}}}=\ev*{c^\dagger_{\vb*{k}}c_{\vb*{k}}}$, averaged over these states, remains close to $2/3$ throughout the BZ [Fig.~\ref{fig:2}(b)].
In the LLL, the Haldane $V_1$ pseudopotential stabilizes the $\nu=1/3$ Laughlin state and its particle-hole conjugate at $\nu=2/3$. The approximate generalized LLL character of the tMoTe$_2$ Bloch states, together with the $V_1$-induced bandwidth reduction near $\nu=1$, thus favors the $\nu=2/3$ FCI above a critical $V_1$.

At the other commensurate filling $\nu=3/4$, we find a TCDW for $V_1\geq3.27$~meV. At $N_s=36$ and $V_1=3.5$~meV, the ED spectrum exhibits four nearly degenerate ground states at the BZ center $\gamma$ and the three edge midpoints $m_{1,2,3}$, separated by a gap from the rest of the spectrum [Fig.~\ref{fig:2}(c)]. The band-projected structure factor $\bar{S}(\vb*{q}) = \ev*{\bar{n}(-\vb*{q})\,\bar{n}(\vb*{q})}/N - \delta_{\vb*{q},0} N$, averaged over the four states, peaks at  $m_{1,2,3}$  [Fig.~\ref{fig:2}(d)].
Both features indicate a charge density wave. Moreover, the many-body Chern number under twisted boundary conditions~\cite{Niu1985quantized} is $1$, identical to that at full filling $\nu=1$, establishing the topological nature of the TCDW \cite{Sheng2024quantum, Liu2025Broken}. The same TCDW is consistently found for the smaller cluster with $N_s=28$.

To probe potential superconducting instabilities, we further calculate the pair-binding energy
\begin{equation}
E_b(N) = E(N+2) + E(N) - 2E(N+1),
\end{equation}
where $E(N)$ is the ground-state energy of the $N$-particle system. A negative $E_b$ indicates that two added particles favor binding over remaining unbound, serving as one diagnostic of pairing in finite systems \cite{White1992pair,Guerci2025from}.

Remarkably, pair binding ($E_b<0$) occurs across all four system sizes ($N_s=31$, 36, 37, and 39) over the filling range $\nu=25/36\approx0.69$ to $\nu=30/39\approx0.77$, with the upper $V_1$ threshold depending on $\nu$ and reaching about $3$~meV [Fig.~\ref{fig:1}(c)]. For $N_s=36$ at $\nu=3/4$, $E_b$ remains negative up to $V_1=2.6$~meV, whereas the TCDW emerges for $V_1\geq3.27$~meV. 
Thus, the pair-binding region is bordered by the Fermi liquid at $V_1=0$, the FCI at $\nu=2/3$, and the TCDW at $\nu=3/4$ and stronger $V_1$.

\textit{Chiral superconductivity.}---
We present several numerical probes of superconducting pairing tendencies in states with $E_b<0$, focusing on $\nu=3/4$, $V_1=1$~meV, and $N_s=36$ unless stated otherwise. At this representative point, the ED spectrum exhibits a nondegenerate ground state at zero momentum, separated from all excited states  [Fig.~\ref{fig:3}(a)]. 
The momentum distribution $\ev*{n_{\vb*{k}}}$ in the ground state reveals a Fermi-surface-like structure around $\gamma$ [Fig.~\ref{fig:3}(b)], consistent with the maximum of the $\nu=1$ Hartree--Fock band at $V_1=1$~meV  [Fig.~\ref{fig:1}(a)].

The momentum-space structure around $\gamma$ [Fig.~\ref{fig:3}(b)] suggests an even-odd effect in pairing: holes at $\vb*{k}$ and $-\vb*{k}$ pair, while the self-inverse momentum $\gamma$ remains unpaired. The paired ground state should contain one hole at $\gamma$ and the remaining holes in pairs, favoring binding for odd hole numbers, $N_h=N_s-N$~\cite{Guerci2025from}. This even-odd effect appears across four system sizes in Fig.~\ref{fig:1}(c) and is further illustrated for $N_s=36$ and 37 in Fig.~\ref{fig:3}(c), where $E_b<0$ only for odd $N_h$.

We next examine the superfluid stiffness $D_s$ to assess phase coherence. We define
$D_s = (\sqrt{3}/8)\,\partial^2 E_g/\partial\Phi_1^2|_{\Phi_1=0}$,
where $E_g$ is the lowest energy as a function of the boundary twist phase $\Phi_1$~\cite{Scalapino1993insulator,Wu2026exact,Ahn2026superconductivity}. As shown in Fig.~\ref{fig:3}(d), $D_s$ remains positive throughout the pair-binding region but decreases monotonically with $V_1$, approaching zero near the upper threshold where $E_b$ also vanishes. This decrease is consistent with the narrowing of the Hartree--Fock band at $\nu=1$ as $V_1$ increases, which implies a larger effective mass. The stiffness curves for $N/N_s=27/36$, $28/37$, and $30/39$ nearly coincide, suggesting that $D_s$ is close to converged with system size. Meanwhile, $-E_b$ reaches maximum at intermediate $V_1$.

The evolution of the low-energy spectrum in Fig.~\ref{fig:3}(e) further reveals how the pair-binding regime gives way to the TCDW. At $V_1=0$, the Fermi liquid exhibits multiple low-energy states. As $V_1$ increases, a nondegenerate ground state with negative $E_b$ emerges, and the charge-neutral gap follows a trend similar to $|E_b|$, first increasing and then decreasing to zero at $V_1\approx2.6$~meV. For $2.7~\text{meV}\lesssim V_1\lesssim3.3$~meV, the low-energy spectrum becomes increasingly dense. At $V_1\approx3.3$~meV, the four lowest levels at $\gamma$ and $m_{1,2,3}$ separate from higher levels, signaling the onset of the TCDW identified in Fig.~\ref{fig:2}(c). The TCDW neutral gap then opens gradually, reaching $0.28$~meV at $V_1=4$~meV.

To characterize the pairing symmetry, we study the pairing amplitude $F(\vb*{k})=\mel{\Psi_{N+2}}{c^\dagger_{\vb*{k}}c^\dagger_{-\vb*{k}}}{\Psi_N}$, where $\Psi_N$ denotes the $N$-particle ground state. For $N_s=36$, both $\Psi_{27}$ and $\Psi_{29}$ are nondegenerate ground states at zero momentum, ensuring that $F(\vb*{k})$ is well defined.
By fermionic antisymmetry, $F(\vb*{k})$ is odd in $\vb*{k}$ and thus vanishes at $\gamma$ and $m_{1,2,3}$. Here $F(\vb*{k})$ is gauge dependent. We fix the gauge by requiring the overlap of the Bloch state with the trial orbital $g(\vb*{r})=(\delta(\vb*{r}),-\delta(\vb*{r}))^{\mathsf T}$, whose components correspond to the two layers, to be real and positive at each momentum except $\gamma$. In this gauge, the threefold rotation symmetry acts trivially for $\vb*{k} \neq \vb*{0}$  as $\hat{C}_{3}c^\dagger_{\vb*{k}}\hat{C}_{3}^{-1}=c^\dagger_{R_{3}\vb*{k}}$, where $R_{3}$ denotes a  counterclockwise rotation by $2\pi/3$. 
As shown in Fig.~\ref{fig:3}(f), $|F(\vb*{k})|$ is largest where $\ev*{n_{\vb*{k}}}(1-\ev*{n_{\vb*{k}}})$ is large, consistent with the Fermi-surface-like structure in Fig.~\ref{fig:3}(b). Moreover, $F(R_{3} \vb*{k})=e^{i 2\pi/3}F(\vb*{k})$, indicating the pairing as chiral $p$-wave.

The ground-state two-particle reduced density matrix (2-RDM) provides an independent characterization of the pairing. Restricting the 2-RDM to zero total momentum, we diagonalize
$D_{\vb*{k}\vb*{k}'}=\ev*{c^\dagger_{\vb*{k}}c^\dagger_{-\vb*{k}}c_{-\vb*{k}'}c_{\vb*{k}'}}$.
The leading eigenvector $\chi_0(\vb*{k})$ identifies the dominant pairing channel~\cite{Yang1962concept}. Its magnitude and phase pattern, shown in Fig.~\ref{fig:3}(g), match those of $F(\vb*{k})$.

To further test the pairing tendency and symmetry, we construct an attractive interaction with the desired pairing form factor $f(\vb*{k})$, 
\begin{equation}
\begin{aligned}
H_{2} &= -U\, b^\dagger b, \qquad b = \frac{1}{\sqrt{2 N_s} } \sum_{\vb*{k}} f(\vb*{k})\, c_{-\vb*{k}} c_{\vb*{k}},\\
f(\vb*{k}) &= \frac{1}{\sqrt{6}} \sum_{n=0}^{5} e^{i  n\pi/3}\, e^{-i \vb*{k}\cdot\vb*{R}_n},
\end{aligned}
\end{equation}
where $U>0$ and $\vb*{R}_n= a_M [-\sin(n\pi/3),\cos(n\pi/3)]$. Here $f(\vb*{k})$ has the same $p$-wave symmetry as $F(\vb*{k})$. Along the interpolation
$H(s) = H_0 + (1-s)H_1 + sH_2$,
which connects the original Hamiltonian $H$ at $s=0$ to the attractive model $H_0+H_2$ at $s=1$, the ground state remains nondegenerate and separated from all excited states by a finite gap, as shown in Fig.~\ref{fig:3}(h). This establishes that the ground state of $H$ is adiabatically connected to a superconducting state with chiral $p$-wave pairing.

\begin{figure}
\centering
\includegraphics[width=\columnwidth]{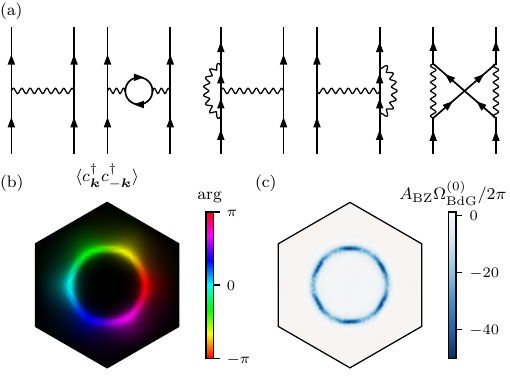}
\caption{(a) Feynman diagrams of the KL pairing interaction to second order in the bare interaction (wavy lines), including the first-order term and the four second-order terms. (b) $\ev*{c^\dagger_{\vb*{k}} c^\dagger_{-\vb*{k}}}$ in the BZ. Hue represents the phase and brightness the magnitude. (c) Berry curvature $\Omega^{(0)}_{\mathrm{BdG}}$ in units of $2\pi/A_{\mathrm{BZ}}$.}
\label{fig:4}
\end{figure}

\textit{Kohn--Luttinger theory.}---
The ED results suggest that repulsion-induced SC emerges from a Fermi liquid at weak $V_1$, motivating a second-order KL analysis~\cite{Kohn1965new}. The effective KL pairing interaction $V_{\vb*{k}\vb*{k}'}^{(s)}$ includes the bare interaction and four second-order diagrams [Fig.~\ref{fig:4}(a)]. The gap  function is defined as $\Delta_{\vb*{k}} = A^{-1} \sum_{\vb*{k}'} V_{\vb*{k}\vb*{k}'}^{(s)} \ev*{c_{-\vb*{k}'} c_{\vb*{k}'}}$. The corresponding mean-field Bogoliubov--de Gennes (BdG) Hamiltonian in the Nambu basis $(c_{\vb*{k}}, c^\dagger_{-\vb*{k}})$ is
\begin{equation}
h_{\vb*{k}} = \begin{pmatrix} \xi_{\vb*{k}} & \Delta_{\vb*{k}} \\ \Delta^*_{\vb*{k}} & -\xi_{-\vb*{k}} \end{pmatrix},
\end{equation}
where $\xi_{\vb*{k}} = \varepsilon_{\vb*{k}} - \mu$, with the chemical potential $\mu$ fixed by the filling. The self-consistent gap equation is
\begin{equation}
\Delta_{\vb*{k}} = \frac{1}{A} \sum_{\vb*{k}'} V_{\vb*{k}\vb*{k}'}^{(s)}\, \frac{f(E^+_{\vb*{k}'}) - f(E^-_{\vb*{k}'})}{E^+_{\vb*{k}'} - E^-_{\vb*{k}'}}\, \Delta_{\vb*{k}'},
\end{equation}
where $f$ is the Fermi function and $E^\pm_{\vb*{k}}$ are the two eigenvalues of $h_{\vb*{k}}$. 
Solving the gap equation at $\nu=0.75$ and $V_1=1$~meV yields a nondegenerate leading pairing channel with critical temperature $T_c=2.44$~K ($k_BT_c=0.21$~meV), comparable to $|E_b|$ from ED. Near zero temperature, the pairing amplitude $\ev*{c^\dagger_{\vb*{k}}c^\dagger_{-\vb*{k}}}$ peaks at the normal-state Fermi surface [Fig.~\ref{fig:4}(b)] and exhibits $p$-wave pairing with the same chirality as $F(\vb*{k})$ from ED.

The mean-field theory also determines the BdG band topology. Let $\psi_{\vb*{k},n}$ and $C^{(n)}$ denote the Bloch wave function and Chern number of the $n$th normal-state band of $\mathcal{H}$. The occupied BdG band derived from the active band $n=0$ has the wave function $\phi_{\vb*{k},0}(\vb*{r}) = \bigl(u_{\vb*{k}}\,\psi_{\vb*{k},0}(\vb*{r}),\, v_{\vb*{k}}\,\psi^{*}_{-\vb*{k},0}(\vb*{r})\bigr)^{\mathsf T}$, where $(u_{\vb*{k}},v_{\vb*{k}})^{\mathsf T}$ is the lower-energy eigenvector of $h_{\vb*{k}}$. Its Berry curvature $\Omega^{(0)}_{\mathrm{BdG}}$, shown in Fig.~\ref{fig:4}(c), integrates to $C^{(0)}_{\mathrm{BdG}}=-2$.

The remaining bands $n\geq1$ are unoccupied in the normal state and remain unpaired in the weak-coupling approximation. Their hole partners, however, lie at negative BdG energies and have wave functions $(0,\psi^{*}_{-\vb*{k},n})^{\mathsf T}$ with Chern numbers $-C^{(n)}$. Since the Chern numbers of all normal-state bands sum to zero, the total BdG Chern number is
\begin{equation}
C_{\mathrm{BdG}}
= C_{\mathrm{BdG}}^{(0)}-\sum_{n\geq1}C^{(n)}
= C_{\mathrm{BdG}}^{(0)}+C^{(0)}
= -1,
\end{equation}
which implies one chiral Majorana edge mode. We emphasize that the BdG topology depends on the full wave functions, incorporating both the normal-state Bloch states and the particle-hole amplitudes, and on the bands away from the Fermi level. 
The sign of $C_{\mathrm{BdG}}$ is opposite to that of the normal-state Chern number $C^{(0)}$ at full filling ($\nu=1$),  similar to  related studies~\cite{MayMann2026how,Tavakol2026pairing,Wang2025chiral}.

\textit{Discussion.}---
In summary, we demonstrate repulsion-driven SC
in the moir\'e band of tMoTe$_2$,
proximate to an FCI at $\nu=2/3$ and a TCDW at $\nu=3/4$, which establishes a filling--interaction phase diagram resolving
all three phases in a realistic moir\'e Chern band.
In the weak-repulsion regime, KL theory identifies a superconducting
Fermi-surface instability with pairing symmetry consistent with ED.

At stronger repulsion, SC gives way to a TCDW
at $\nu=3/4$, separated by a nearly gapless regime
[Fig.~\ref{fig:3}(e)].
At intermediate repulsion, the same $V_1$ supports an FCI
at $\nu=2/3$ and SC over
$0.7\lesssim\nu\lesssim0.77$.
Between the two phases, we find low-energy bands of fractionally
charged anyons, with one- and two-anyon bands at
$(N,N_s)=(23,34)$ and $(24,35)$, respectively.
Their connection to SC remains an open question.

Comparison with experiment \cite{Xu2025signatures} is encouraging.
Our model captures the FCI at $\nu=2/3$ and predicts a superconducting filling range consistent with
the experimental interval $0.71\lesssim\nu\lesssim0.76$.
While the experiment reports the reentrant integer Chern insulator near
$\nu=0.7$ under a finite displacement field, our
zero-displacement-field model yields a TCDW at $\nu=3/4$.
Thus, a single microscopic model with only one tuning 
parameter $V_1$ captures several key correlated phases.

Our theory predicts that reducing $V_1$ through
dielectric screening destabilizes the FCI before SC.
Moreover, the superconducting BdG Chern number
$C_{\mathrm{BdG}}=-1$ is opposite to the normal-state
Chern number $C^{(0)}=1$ at $\nu=1$,
a chirality reversal that could be probed by comparing
thermal Hall and electrical Hall transport.

\begin{acknowledgments}
\textit{Acknowledgments.}---
We thank Tingxin Li for valuable discussions. 
This work  was supported by the National Key Research and Development Program of China (Grants No. 2021YFA1401300 and No. 2022YFA1402400), and the National Natural Science Foundation of China (Grants No. 12274333 and No. 12550404).  
\end{acknowledgments}

\bibliography{references}

\end{document}